# DEEP LEARNING FOR EEG SEIZURE DETECTION IN PRETERM INFANTS

**Alison O'Shea***
*Irish Centre for Maternal and Child Health Research (INFANT),*
*Department of Electrical and Electronic Engineering,*
*University College Cork, Cork, Ireland*
*Email: alisonoshea@umail.ucc.ie*

**Rehan Ahmed**
*Irish Centre for Maternal and Child Health Research (INFANT),*
*Department of Electrical and Electronic Engineering,*
*University College Cork, Cork, Ireland*

**Gordon Lightbody**
*Irish Centre for Maternal and Child Health Research (INFANT),*
*Department of Electrical and Electronic Engineering,*
*University College Cork, Cork, Ireland*

**Sean Mathieson**
*Irish Centre for Maternal and Child Health Research (INFANT),*
*Department of Pediatrics and Child Health,*
*University College Cork, Cork, Ireland*

**Elena Pavlidis**
*Irish Centre for Maternal and Child Health Research (INFANT),*
*University College Cork, Cork, Ireland*
*Child Neuropsychiatric Unit, Medicine and Surgery Department,*
*University of Parma, Italy*

**Rhodri Lloyd**
*Irish Centre for Maternal and Child Health Research (INFANT),*
*University College Cork, Cork, Ireland*

**Francesco Pisani**
*Child Neuropsychiatric Unit, Medicine and Surgery Department,*
*University of Parma, Italy*

**Willian Marnane**
*Irish Centre for Maternal and Child Health Research (INFANT),*
*Department of Electrical and Electronic Engineering,*
*University College Cork, Cork, Ireland*

**Geraldine Boylan**
*Irish Centre for Maternal and Child Health Research (INFANT),*
*Department of Pediatrics and Child Health,*
*University College Cork, Cork, Ireland*

**Andriy Temko**
*Irish Centre for Maternal and Child Health Research (INFANT),*
*Department of Electrical and Electronic Engineering,*

EEG is the gold standard for seizure detection in the newborn infant, but EEG interpretation in the preterm group is particularly challenging; trained experts are scarce and the task of interpreting EEG in real-time is arduous. Preterm infants are reported to have a higher incidence of seizures compared to term infants. Preterm EEG morphology differs from that of term infants, which implies that seizure detection algorithms trained on term EEG may not be appropriate. The task of developing preterm specific algorithms becomes extra-challenging given the limited amount of annotated preterm EEG data available.

This paper explores novel deep learning architectures for the task of neonatal seizure detection in preterm infants. The study tests and compares several approaches to address the problem: training on data from full term infants; training on data from preterm infants; training on age-specific preterm data and transfer learning. The system performance is assessed on a large database of continuous EEG recordings of 575 hours in duration.

It is shown that the accuracy of a validated term-trained EEG seizure detection algorithm, based on a support vector machine classifier, when tested on preterm infants falls well short of the performance achieved for full term infants. An AUC of 88.3% was obtained when tested on preterm EEG as compared to 96.6% obtained when tested on term EEG. When re-trained on preterm EEG, the performance marginally increases to 89.7%. An alternative deep learning approach shows a more stable trend when tested on the preterm cohort, starting with an AUC of 93.3% for the term-trained algorithm and reaching 95.0% by transfer learning from the term model using available preterm data. The proposed deep learning approach avoids time-consuming explicit feature engineering and leverages the existence of the term seizure detection model, resulting in accurate predictions with a minimum amount of annotated preterm data.

## 1. Introduction

Seizures are the primary biomarker of neurological dysfunction in term and preterm infants with an incidence of 2-3 per 1000 live births in term infants, and much higher incidences, ranging from 5-30 per 1000 live births, reported in preterm infants.[1-3] It is critical to detect and treat seizures at the earliest opportunity. The clinical diagnosis of seizures in infants is challenging without brain monitoring due to the prevalence of electrographic-only seizures which show no visible clinical manifestations.[4,5] The identification of preterm seizures with clinical manifestations is complicated by the vast repertoire of jerky movements that are often seen in preterms, many of which are essential for normal sensorimotor development.[6,7] These jerky movements can be difficult to distinguish from seizures which can result in unnecessary treatment of infants with anti-epileptic drugs.

Continuous EEG monitoring is the optimal method available for the detection of seizures in infants. Interpretation of neonatal EEG requires highly trained healthcare professionals and as a result, it is limited to specialized centers. Several computer methods have been developed for the detection of seizures in term infants.[8-17] Over the past 20 years term EEG seizure detection algorithms (T-SDAs) have passed several stages of development.[18] The early algorithms relied on signal processing routines to extract information from the EEG followed by hand-tuned rule-threshold decision making. Over many years of research simple features have evolved into more sophisticated hand-crafted characteristics while the rules and thresholds have transformed into data-driven classifier-based decision making. State-of-the-art results have been reported by a recent deep learning (DL) solution which combines feature extraction and classification into a single end-to-end optimization problem.[19] DL has also shown success in detecting seizures in adult EEG.[20]

While there has been research in the area of automated classification of patterns in preterm background EEG,[21-25] no studies on the automated detection of seizures in preterm EEG exist. There are several important differences between term and preterm EEG which makes the development of preterm EEG seizure detection algorithms (P-SDA) challenging. One challenging aspect to consider when developing a P-SDA is data availability. Recording multichannel EEG in the preterm is difficult due to the clinical condition of preterm infants and physical limitations such as the small head circumference. Specially trained staff are required to record preterm EEG with the correct equipment, and careful preparation.[26] Good quality seizure and non-seizure data sets, which are required to train, test and validate an algorithm for automated analysis, are therefore very difficult to obtain. The algorithmic solution to this problem must consider that comparatively small datasets will be available.

The characteristics of preterm EEG are very different to those of the term infant; this is another aspect to consider when designing a P-SDA. The features of preterm EEG that make P-SDA development more involved include the intermittent background pattern (tracé discontinu) which consists of periods of low-voltage activity (inter-burst intervals) alternating with periods of short duration higher-voltage activity, known as bursts or spontaneous activity transients. Additionally, preterm seizures are usually focal/regional with less propagation than seizures of term neonates due to ongoing synaptogenesis and a relative paucity of inter-regional connectivity.[27] The frequency of most preterm seizures lies in the delta band, but has been shown to evolve with age, and seizure morphology tends towards repetitive sharp waves, rather than spike and wave pattern.[27,28] Preterm seizures are usually shorter in duration and demonstrate less evolution in terms of frequency and morphology during an ictal event. Figure 1 (a) shows an example of a preterm seizure with little spatial or frequency evolution. Given the complexity of preterm EEG, the set of seizure-descriptive features which were developed for the T-SDA machine learning model may not be optimal for preterm seizure detection.

Preterm EEG evolves rapidly as the infant matures and a thorough knowledge of the age appropriate maturational features of the EEG is required in order to identify seizures accurately.[29] Preterm EEG patterns change throughout gestation particularly in terms of voltage and length of discontinuous periods. Bursts of activity lengthen and inter-burst intervals progressively shorten during the preterm period and it has been suggested that this evolution from discontinuity to continuity represents a transition from primarily endogenously generated sporadic cortical activities in preterm to continuous sensory driven activity towards term.[30] In addition, specific features appear and disappear at different gestational ages (GA) and features such as asynchrony between hemispheres can be normal or abnormal depending on the GA.[29,31] There are several intermittent age appropriate maturational features such as delta brushes, and transient theta bursts over the temporal and occipital regions. The frequency content of the EEG also changes with a predominance of slow delta activity at early preterm ages and an increase in faster activities towards term.[32] Figure 1 (b) and (c) show examples of preterm background EEG patterns. The patterns in Figure 1 (b) and (c) show the characteristic changes that occur with GA. The question arises as to whether there is a need to employ models that are specific to GA, due to the evolution of preterm characteristics with GA; this work will look at both generalized and age specific classifier development. Given the scarcity of preterm data, it is important to consider efficient data usage when developing age specific preterm models, which are each naturally supported by a much smaller subset of the overall training data.

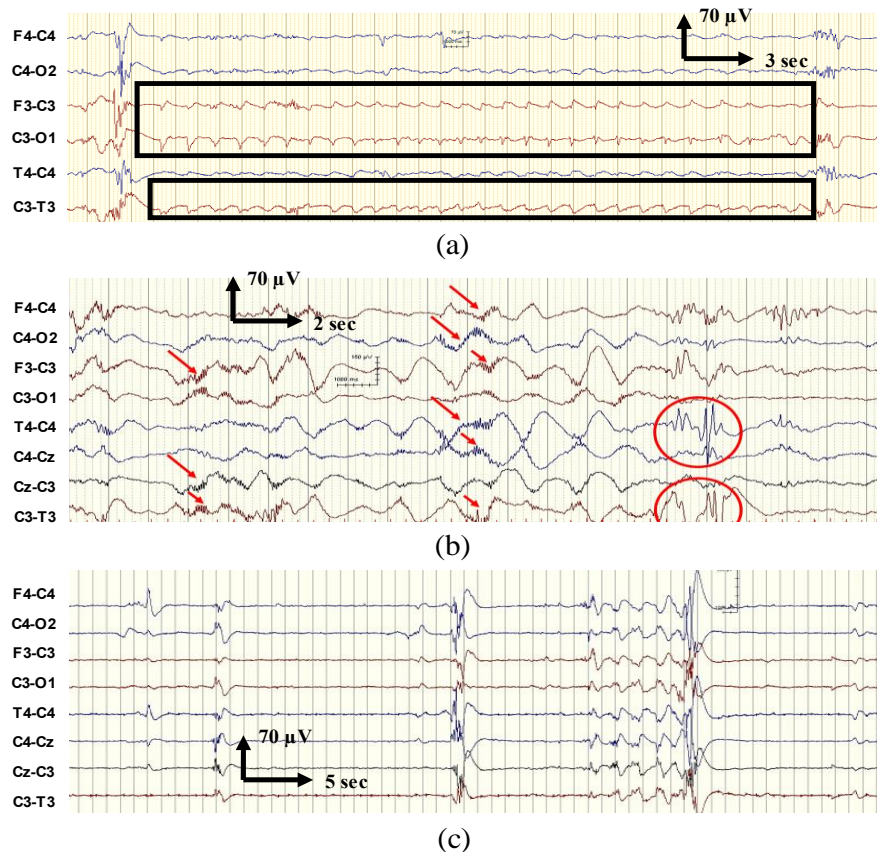

Fig. 1. Examples of preterm background EEG patterns and seizure patterns. (a) A representative example of preterm seizure showing a short 30-second lateralized seizure with rhythmic delta morphology and little evolution at 28 weeks GA, (b) runs of background semi-rhythmic delta activity with delta brush (arrows) and premature temporal theta transients (circles) at 31+5 weeks GA, (c) discontinuity at 24 weeks GA.

Previous work in the area of epileptic seizure detection in adult EEG has already shown that the algorithms developed for the adult population are not suitable for term neonatal EEG.[12, 33-35] The clinical urgency and the outlined differences between term and preterm EEG need to be accommodated in the choice of data partition (train and test), system design, at the level of feature engineering and predictive modelling.

This study poses and addresses the following research questions:
  i. Can the algorithms which were designed to detect seizures in term neonatal EEG also be used to detect seizures in preterm EEG?
  ii. What are the steps to improve the model performance on preterm EEG?
  iii. How can the existence of the term-EEG seizure detection model and an abundance of term EEG data be leveraged to create an accurate preterm EEG seizure detection system with minimum preterm annotated EEG data in the minimum amount of time?

This study is organized as follows: section 2 provides an overview of the previously developed T-SDAs and presents the proposed P-SDAs along with the experimental setup; section 3 presents details of the databases (DB) used; section 4 presents the experimental results which are then discussed in section 5.

## 2. Methods

This work explores two baseline T-SDAs which were trained on a term EEG dataset, namely the SVM based T-SDA[16] and a deep-learning based T-SDA.[19] These two algorithms were initially assessed on the preterm test DB (the Cork dataset of long continuous preterm EEG recordings) to establish the baseline preterm performance. Subsequently, the algorithms were retrained using preterm EEG from the preterm training DB and re-assessed on the preterm test DB. The deep learning algorithm retraining involved several modifications. While the full descriptions of the algorithms have been previously published[16,19], the general structure of the algorithms in included here to enable discussion of the performance results.

### 2.1. *SVM T-SDA*

Figure 2 (a) shows the system level representation of the SVM algorithm. This system is divided into four main stages: the preprocessing stage, the feature extraction stage, the SVM classification stage and the postprocessing stage. During preprocessing the multi-channel EEG is down-sampled from either 1024Hz or 256Hz (depending on the recording frequency) to 32Hz with an anti-aliasing filter set at 12.8Hz and a high pass filter set to 0.5Hz. The signal is then segmented into single channel 8-second windows with a 4-second overlap. A set of 55 features are extracted from these 8-second windows; these features characterize the EEG activity in the time domain, the frequency domain and using information theory measures. These features have been engineered specifically for term EEG and validated in a number of previous studies on term EEG, such as neonatal seizure detection,[16,36,37] grading term background EEG,[38] neurological outcome prediction in term infants[39] and even adult seizure detection and prediction.[33,40] The extracted feature vectors are then fed to the SVM classification stage of the system where the inputs are assigned a probability of being a seizure. This probability is smoothed and postprocessed. Temko et al.[16,41] have reported in detail on this algorithm and on the set of 55 features.

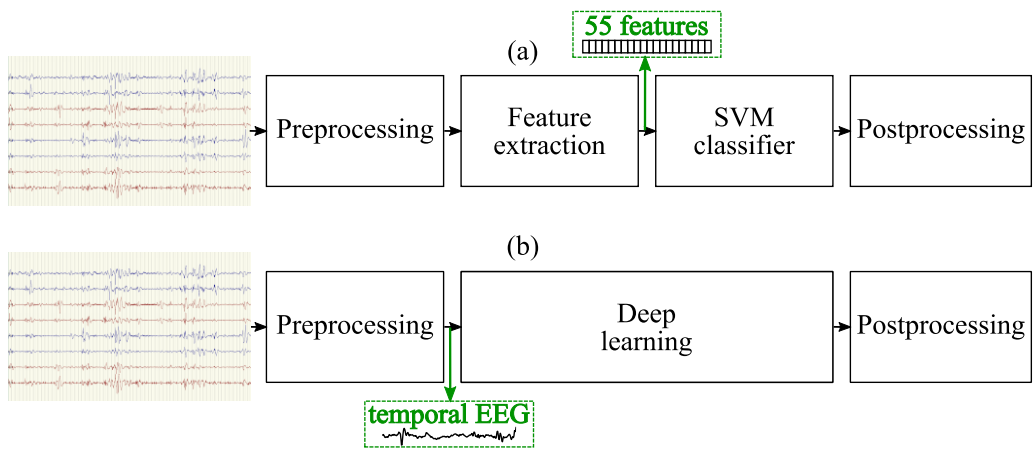

Fig. 2. Comparison of the (a) SVM and (b) DL algorithms. The SVM algorithm relies on hand-crafted engineered features that aim to convey all the information from the raw preprocessed EEG. The features are fed into the SVM and postprocessed to produce a probability of a seizure being present in the given time interval. In the deep learning architecture, the neural network is applied to raw preprocessed EEG directly and requires no distinct feature extraction stage.

## 2.2. DL T-SDA

Figure 2 (b) shows the system level representation of the deep learning algorithm. The deep learning system utilizes the same preprocessing and postprocessing stages as the SVM algorithm, but the feature extraction and classification stages are combined into one end-to-end deep learning network. The algorithm is applied to temporal multi-channel EEG signals directly; this EEG has been down-sampled to 32Hz (anti-aliasing filter set to 12.8Hz and a high pass filter set to 0.5Hz) and split into 8-second windows with a 7-second overlap.

Convolutional filters are used to extract data-driven information from the 8-second multi-channel windows of EEG. A series of hierarchical non-linear filtering operations extract a complex internal representation of the EEG, with the final convolutional layers performing the classification, as shown in Table 1. The designed fully-convolutional architecture was optimized on term EEG.[19]

An important aspect of this architecture is that it can be trained on multi-channel EEG signals - therefore the seizure annotations do not need to be channel specific. EEG windows of (N,256,1) dimensions are input to the input convolutional layer of the network; N is the number of EEG channels and 256 represents 8-second windows of EEG sampled at 32Hz. All convolutional filters in the network are applied across the temporal dimension only, so N, the number of EEG channels remains unchanged. In the penultimate layer, the average of all feature maps in the temporal dimension is calculated, then the max across all feature maps in the EEG channel dimension is calculated. This operation is analogous to the max operator which is applied across EEG channels in the SVM T-SDA postprocessing routine.

The architecture takes multi-channel EEG inputs and produces a single probabilistic output. This property meant that the T-SDA algorithm developed by O'Shea et al.[19] was trained on a dataset of over 800 hours EEG. In contrast to this, the SVM T-SDA was trained on a subset of around 2% of the same term EEG dataset because it requires channel-specific seizure annotations during the training stage. The generation of channel-specific seizure annotations is a laborious process and typically only a subset of seizure events in a dataset are annotated in this way.

The model was trained using a categorical cross-entropy loss function and a learning rate of 0.01. Three babies from the training set were removed and utilized as a patient independent validation set for early stopping. The Area Under the receiver operating Curve (AUC) score on the validation set was calculated after each epoch - if the validation AUC score did not improve for 8 training iterations the training was stopped and the network which resulted in the best validation

Table 1. The structure of each feature extraction block (a) and classification block (b) used in the deep learning T-SDA. The full convolutional DL algorithm consists of 3 feature extraction blocks (a) followed by one classification block (b). This gives a total convolutional depth of 10 layers.

| (a) Architecture of feature extraction building block | | | | |
|---|---|---|---|---|
| Layer Name | Feature Maps | Size | Stride | Activation |
| Convolution | 32 | 3 | 1 | ReLU |
| Convolution | 32 | 3 | 1 | ReLU |
| Convolution | 32 | 3 | 1 | ReLU |
| Batch Normalization | | | | |
| Average Pooling | | 4 | 3 | |

| (b) Architecture of classification building block | | | | |
|---|---|---|---|---|
| Layer Name | Feature Maps | Size | Stride | Activation |
| Convolution | 2 | 3 | 1 | ReLU |
| Global Pooling | - | - | - | |
| Output | - | - | - | Softmax |

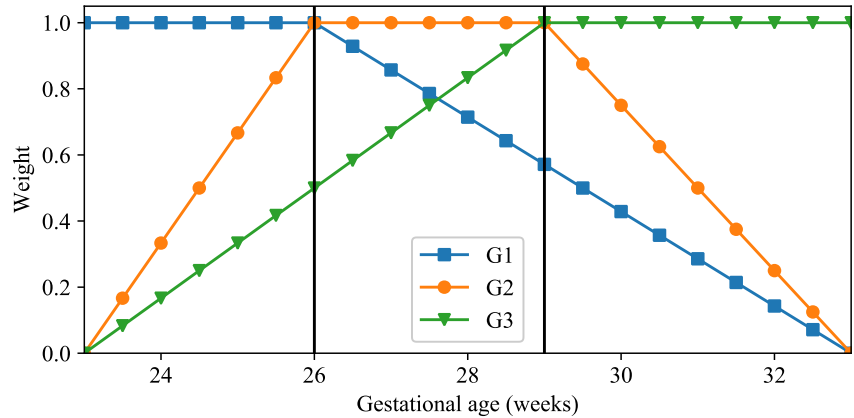

Fig. 3. The weighting applied to each preterm infant's GA during training of the GA specific DL classifier. These weights are utilized during the training of the GA specific algorithm to ensure that all of the training data was used despite the developed classifiers being age-specific.

AUC score was selected as the final network. This routine was repeated three times, randomly selecting a different set of babies for the validation set on each iteration. This resulted in three separate models, each based on a slightly different set of training babies. In order to fully utilize all three models, they are combined using an ensembling technique. The average of the probabilistic outputs from the three trained networks was taken to be the final probabilistic output at the inference stage.

**2.3. *Development of accurate P-SDAs***

The simplest way to build a P-SDA is to re-use an existing T-SDA algorithm but retrain it on preterm EEG data. In this manner, the same pre-processing, feature extraction, modelling and post-processing stages can be employed. In the first experiment both T-SDAs were trained using the preterm training data (Parma dataset) and were then tested on the preterm test data (Cork dataset). This experiment will show how suitable the system choices (engineered features, DL architecture, etc.) which were made for term EEG scenario are for classification of preterm EEG. For the DL P-SDA three separate models were trained, each with a different random weight initialization and each with a different randomly chosen subset of held-out training infants for early stopping. These models were combined by taking the average probabilistic outputs to generate one DL P-SDA model.

It is known that the GA of a preterm has a large effect on EEG morphology.[29] A second experiment was therefore proposed to determine if the use of age-specific models improves the performance of the preterm SDAs. In this experiment, the training and test datasets were divided into three subgroups according to the gestational ages of the infants: Group 1 (GA < 26 weeks), Group 2 (26 weeks ≤ GA ≤ 29 weeks) and Group 3 (GA > 29 weeks). The system was then trained and tested using only the data of the chosen subgroup. This results in three different models with each model covering a specific GA range.

Given the scarcity of the training data, segmenting data into GA subgroups comes at a cost of reducing the training data for each model. The SVM is known to perform well with limited amount of training data and thus for the SVM-based SDAs the training process for GA-specific experiment is unchanged. However, the DL is known to require large amounts of training data to be accurate, and the scarcity of available training data in GA-specific experiment poses a real problem for the DL-based SDAs. To overcome the limited training data availability when training the GA-specific DL SDAs the training process has been modified in three ways:

i) The DL GA-specific P-SDA was trained using transfer learning instead of retraining from scratch. In particular, the network was initialized with weights from one of three T-SDA algorithms which were trained on large amounts of term EEG. The data from each GA group were used to fine-tune the trainable network parameters. All layers in the fully convolutional architecture are updated during the training process.

ii) Layer-wise adaptive learning rate scaling mechanism (LARS) was utilised.[42] LARS extends stochastic gradient descent with momentum to determine a learning rate per layer. This is done to account for the potential variation of parameter (weights and biases) and gradient magnitudes in each layer during the transfer learning stage. In typical standard gradient descent updates, the weight update is $w_{t+1}^i = w_t^i - \eta \nabla w_t^i$, where $w_t^i$ is the weight in the $i^{th}$ layer at step $t$, $\nabla w_t^i$ is the gradient of the weight with respect to the loss function and $\eta$ is the learning rate. LARS scales the weight update by the layer-wise ratio of the $l_2$-norm of the parameters and the $l_2$-norm of the gradients.

$$w_{t+1}^i = w_t^i - \eta \frac{\|w\|}{\|\nabla w\|} \nabla w_t^i \quad (1)$$

iii) Instead of excluding the training data from other GA groups all the available training data are included in the DL-transfer learning experiment. Each training sample is weighted based on one of the three membership functions shown in Figure 3. In particular, the training samples that belong to the chosen GA group are given a weight of 1.0 in the loss; the remaining data are given a weight which linearly decays depending on the difference between their GA and the GA threshold of the group for which the P-SDA is trained.

The final experiment uses classifier fusion to increase the power of the developed P-SDA. The fusion of classifiers has been proven to improve performance, especially if the classifiers represent diverse ways of approaching the same task.[43] To assess this hypothesis, two simple fusion schemes were tested using either the weighted arithmetic mean (2) or the weighted geometric mean (3) over the probabilistic output of the classifier, P, which are defined as follows, for $\alpha_i \geq 0, \sum_i \alpha_i = 1$:

$$P_{fusion\_a\_mean} = \sum_i \alpha_i P_i \quad (2)$$

$$P_{fusion\_g\_mean} = \prod_i P_i^{\alpha_i} \quad (3)$$

For two classifiers, Eq. 2 can be rewritten as $P_{fusion\_a\_mean} = \alpha P_{T-SDA} + (1-\alpha) P_{T-SDA}$ and Eq. 3 becomes $P_{fusion\_g\_mean} = P_{T-SDA}^{\alpha} P_{P-SDA}^{1-\alpha}$, where $0 \leq \alpha \leq 1$.

### 2.4. Performance metrics

To quantify the performance of the system, the AUC score, which measures sensitivity versus specificity, is used. Sensitivity (4) and specificity (5) are defined as the epoch-wise accuracy of each class (seizure and non-seizure), respectively. Each epoch is labelled as either true positive (TP), false positive (FP), true negative (TN), false negative (FN); sensitivity and specificity are summations of these labels. All AUCs reported in this work are generated by concatenating the probabilistic outputs and the true labels of multiple preterms.

$$Sensitivity = \frac{\sum TP}{\sum TP + \sum FN} \quad (4)$$

$$Specificity = \frac{\sum TN}{\sum TN + \sum FP} \quad (5)$$

Second, we report the event-based clinically relevant metric of the number of correctly detected events at a specific false detection per hour (FD/h) threshold. The seizure detection rate is the percentage of seizure events that are correctly detected. A correct detection is any overlap of a predicted positive seizure label with a seizure event in the ground truth. The FD/h metric indicates the number of times that a clinician will have to check the results of an SDA in vain; this is an unwanted consequence of employing an SDA in the NICU. This metric is clinically relevant because clinical staff will only trust an SDA if the FD/h is at a tolerably low level.[44]

### 3. Databases

Two datasets containing annotated neonatal seizures were used in this work to develop and test the P-SDAs. The datasets were recorded in two separate hospitals (Cork and Parma) and annotated by different neurophysiologists. All recordings were annotated by a single neonatal neurophysiology expert (EP or RL). A seizure on the neonatal EEG was defined as a sudden and evolving repetitive stereotyped waveform with a definite start, middle and end, lasting for at least 10-seconds on at least one EEG channel.[45]

Table 2. Cork dataset of preterm EEG. This dataset consists of long EEG recordings with seizures annotated. The seizures have temporal annotations, but the channel specific location of each seizure event is not available. Corrected age shows the age of first seizure event, this is used as the measure of GA for each infant.

| Corrected age (Weeks) | Record Length (hours) | Seizure events | Seizure duration | | Seizure statistics | | |
|---|---|---|---|---|---|---|---|
| | | | >1min | <1min | Mean duration | Min duration | Max duration |
| 23 | 24.00 | 0 | - | - | - | - | - |
| 24 | 48.02 | 84 | 24 | 60 | 48'' | 12'' | 5'46'' |
| 24 | 24.02 | 0 | - | - | - | - | - |
| 25 | 58.80 | 6 | 0 | 6 | 25'' | 20'' | 29'' |
| 25 | 99.56 | 49 | 38 | 11 | 2'44'' | 29'' | 5'54'' |
| 25 | 12.02 | 0 | - | - | - | - | - |
| 26 | 12.01 | 0 | - | - | - | - | - |
| 26 | 24.00 | 0 | - | - | - | - | - |
| 26 | 67.10 | 1 | 0 | 1 | 41'' | 41'' | 41'' |
| 26 | 14.76 | 0 | - | - | - | - | - |
| 29 | 48.49 | 15 | 7 | 8 | 51'' | 33'' | 1'20'' |
| 29 | 12.15 | 0 | - | - | - | - | - |
| 29 | 12.09 | 0 | - | - | - | - | - |
| 30 | 24.06 | 0 | - | - | - | - | - |
| 31 | 24.03 | 0 | - | - | - | - | - |
| 31 | 70.13 | 149 | 94 | 55 | 1'26'' | 15'' | 22'32'' |
| Total | 575.24 | 304 | 163 | 141 | | | |

### 3.1. *Cork dataset of preterm EEG: Preterm EEG test DB*

The first dataset was recorded at Cork University Maternity Hospital (CUMH), Ireland. Ethical approval for the collection and analysis of the data was granted by the Clinical Research Ethics Committee of the Cork Teaching Hospitals, Ireland. Written informed parental consent was obtained for each study. Table 2 presents the details of the Cork dataset. The dataset consists of continuous multichannel video-EEG recordings from infants less than 32 weeks gestation.[2]

A modified 10-20 system, neonatal electrode montage was used; 8 channels were selected for analysis: F4-C4; C4-O2; F3-C3; C3-O1; T4-C4; C4-Cz; Cz-C3; C3-T3. Three EEG machines were used for data collection: the NicoletOne EEG system, the NeuroFax EEG-1200, and the CNS-200 EEG and Multimodal Monitor. EEG was recorded at a sampling frequency of either 256Hz or 1024Hz and a reference electrode was placed at Fz. Disposable, single-patient surface electrodes were utilized.

The dataset contains 6 preterms with seizure events and 10 control preterms without seizure events. The addition of non-seizure infants to the testing dataset ensures that the developed algorithms are tested under conditions were high levels of specificity are enforced. It can be seen from Table 1 that over 45% of all seizures in this dataset were short seizures of less than 1-minute in duration. Different seizure morphologies were observed during analysis, including rhythmical delta; rhythmical triphasic sharp waves; sharp and slow waves; spike and slow waves.

The dataset consists of 575 hours of continuous EEG recordings. This dataset was not edited to remove the large variety of artefacts commonly encountered in the real-world neonatal intensive care unit (NICU) environment. Therefore, this dataset is truly representative of the real-life situation in the NICU and allows for a realistic estimate of the preterm seizure detection algorithm's performance. For these reasons, this dataset is used for testing purposes in this work.

### 3.2. *Parma dataset of preterm EEG: Preterm EEG training DB*

The second dataset was collected in Parma University Hospital, Italy. Ethical approval for the collection and analysis of the data was granted by the Ethics Committee of Parma. Written informed parental consent was obtained. Recordings were started, on average, 13 days (range 1-47 days) after birth. Preterms, with confirmed video-EEG seizures, were admitted to the NICU of Parma University Hospital. A sample dataset was then created with the following inclusion criteria: 1) infants born before 32 weeks GA; 2) video-EEG-confirmed neonatal seizures; 3) seizure onset before 36 weeks GA.

Table 3. Parma dataset of preterm EEG. This dataset consists of EEG recordings with relatively short duration, in comparison to the Cork dataset. All seizure events in this dataset have temporal annotations, a number of the seizure events also have channel-specific annotations. Corrected age shows the age of first seizure event.

| Corrected age (Weeks) | Record length (hh:mm:ss) | Seizure events | Seizure duration | |
|---|---|---|---|---|
| | | | > 1-min | < 1-min |
| 24 | 4:49:01 | 26 | 14 | 12 |
| 24 | 1:29:55 | 2 | 2 | 0 |
| 25 | 0:37:23 | 12 | 6 | 6 |
| 26 | 1:48:40 | 8 | 0 | 8 |
| 26 | 0:32:48 | 1 | 1 | 0 |
| 27 | 1:02:02 | 3 | 0 | 3 |
| 27 | 2:00:16 | 12 | 9 | 3 |
| 28 | 1:54:46 | 13 | 1 | 12 |
| 28 | 0:57:23 | 1 | 1 | 0 |
| 29 | 1:31:36 | 2 | 0 | 2 |
| 29 | 1:01:46 | 1 | 1 | 0 |
| 30 | 1:10:04 | 5 | 2 | 3 |
| 30 | 0:38:54 | 1 | 1 | 0 |
| 30 | 1:10:35 | 15 | 8 | 7 |
| 31 | 1:16:29 | 4 | 4 | 0 |
| 32 | 0:51:38 | 21 | 17 | 4 |
| 32 | 0:31:03 | 1 | 1 | 0 |
| **Total** | **23:24:19** | **128** | **68** | **60** |

Table 3 details the Parma dataset. Depending on the infant's head size, electrodes were applied according to the international 10-20 system modified for infants using 14 recording channels. A Micromed 21-channel synchronized video-EEG machine was utilized for EEG recordings. All EEG was recorded at a sampling frequency of 256Hz and a reference electrode was placed at Fz. Disposable, single-patient surface electrodes were utilized. Recordings continued until a complete cycle of wakefulness, quiet, and active sleep was obtained. When state changes were not clearly distinguishable, the recording continued for at least 60-minutes.

If an infant had several recordings in the same day or serial follow-up recordings, only the files with at least one seizure were selected for inclusion in the dataset. The dataset is primarily comprised of short EEG recordings lasting an average of 1 hour and 19 minutes (range: 0.5-4.8 hours). Approximately half of all seizures in this dataset were less than 1-minute in duration. Most seizures had per channel annotations which are usually required in order to train machine learning algorithms. These short recordings are not reflective of the long duration recording practices which neonatal SDA development targets. A recent clinical trial indicated that SDAs had the most benefit to seizure detection rates in NICUs during the unsociable hours of weekends[46] - for this reason, this dataset was used for training purposes only.

### 3.3. *Comparison of data statistics between term and preterm EEG*

Figure 4 compares the duration of seizure events in the preterm datasets utilized in this work against a dataset of EEG recordings with annotated seizures from a population of term infants. The dataset of term infants was utilized to train the SVM T-SDA and the DL T-SDA, as is reported in previous works.[19,40] This term DB consisted of long recordings (>24h duration) from 18 infants who suffered from seizures and short recordings (~1h duration) from 54 infants who experienced hypoxic-ischemic encephalopathy but did not have any confirmed seizure events. All EEG recordings were associated with full term infants between 39w and 42w GA. The dataset totaled over 834h in duration and contained 1389 annotated seizure events.

Table 4. Performance of various SDAs reported on the Term and Preterm test DBs. Performances are reported in AUC (%).

| SDA | Term EEG test DB | Preterm EEG test DB |
|---|---|---|
| SVM T-SDA | 96.6 | 88.3 |
| SVM P-SDA (retrained from scratch) | - | 89.7 |
| | | |
| DL T-SDA | **98.5** | 93.3 |
| DL P-SDA (retrained from scratch) | - | **93.5** |

Figure 4 indicates that preterms have double the proportion of short seizures, approximately 46% (less than 1-minute duration), when compared to the term infants. Moreover, preterm datasets have the smallest proportion of seizures with duration of greater than 5 minutes. While the distributions of seizure durations in the term and preterm populations differ, the seizure event statistics between the two preterm datasets, train DB and test DB, are similar.

## 4. Experimental results

Table 4 shows the results of the various term and preterm SDAs when tested on Preterm EEG test DB. It can be seen that the SVM T-SDA drops its performance from an AUC of 96.6% on the term test DB to an AUC of 88.3% on the preterm test DB, which is equal to an 8.3% absolute drop in performance. For the DL-based SDAs, a smaller decrease in AUC of 5.2% absolute is observed, dropping from an AUC of 98.5% to an AUC of 93.3%. When retrained on the training preterm DB, the performance of both the SVM and DL SDAs marginally improves, with a larger improvement observed for the SVM P-SDA. The SVM P-SDA results in a 1.4% increase in AUC and the DL P-SDA results in a 0.2% increase in performance. Overall, it can be seen the DL SDAs outperformed the SVM SDAs both on term and preterm EEG.

Table 5 reports the GA-group specific results for the SDAs examined in Table 4 and also for P-SDAs that were retrained using the GA-specific data from the preterm training DB. The DL GA-specific P-SDA utilizes all three training modifications which were detailed in Section 2.3. It can be seen from Table 5 that the performance of both the T-SDA and P-SDA differs considerably across the GA groups. While overall results of retraining from scratch show marginal improvement over their corresponding T-SDAs counterparts, the distribution of scores in the three groups between P-SDAs (retrained from scratch) and T-SDAs are very different. When comparing the GA-specific retraining with simple retraining, the SVM P-SDA did not improve in all 3 groups, whereas the GA-specific transfer learning for the DL P-SDA improved across all three groups; improving from an AUC 93.3% to 95.0% when tested across all babies concatenated in the preterm test DB.

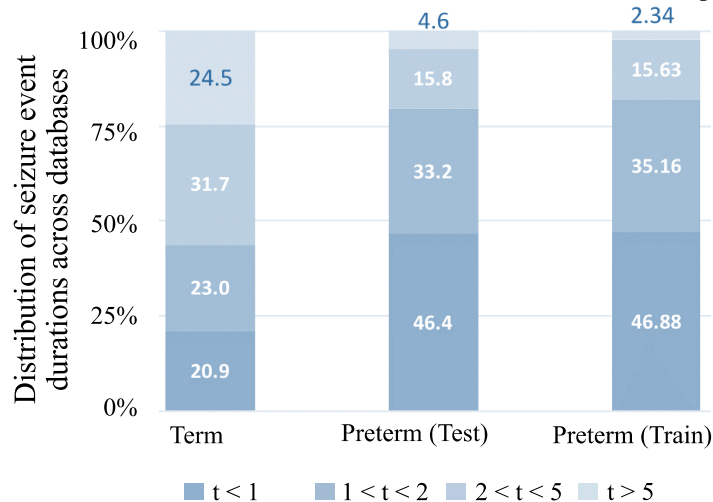

Fig. 4. Distribution of seizure events according to their duration in the preterm datasets and compared to a term dataset. t represents the duration of the seizure event in minutes.

Table 5. GA specific results for the DL and SVM SDAs reported as AUC (%). The classifier performances when tested on each GA group separately are reported alongside the performance of each classifier when tested on the entire test dataset. The T-SDA and P-SDA (retrained from scratch) both for SVM and DL cases have the same single model as in Table 4. For the GA specific P-SDA, three separate algorithms were trained using only data from their corresponding GA group in the preterm training DB for SVM SDAs or using modifications explained in Section 3.3 for DL SDAs. The best performances in column are shown in bold.

| **SDA** | **Group 1** | **Group 2** | **Group 3** | **Preterm EEG test DB** |
|---|---|---|---|---|
| SVM T-SDA | 93.6 | 87.1 | 86.8 | 88.3 |
| SVM P-SDA (retrained from scratch) | 87.1 | 71.8 | **96.2** | 89.7 |
| SVM P-SDA (GA-specific, from scratch) | 82.6 | 91.6 | 90.7 | 78.5 |
|  |  |  |  |  |
| DL T-SDA | 91.8 | **96.0** | 93.8 | 93.3 |
| DL P-SDA (retrained from scratch) | 94.4 | 84.3 | 92.1 | 93.5 |
| DL P-SDA (GA-specific, transfer learning) | **95.8** | 93.8 | 93.4 | **95.0** |

Figure 5 shows the results of clinical metrics, seizure detection rate vs FD/h, for the best combination of the SDA systems, and the results of the individual SDA systems used in the ensemble (DL T-SDA, DL P-SDA GA-specific transfer). The method of fusion and the best weighting of each classifier were found on the validation data (previously utilised for early stopping in DL systems). Figure 6 shows the AUC obtained on the validation set (a subset of the training data) as a function of the chosen weight for both geometric and arithmetic mean ensembling methods. It can be seen from Figure 6 that the geometric mean of probabilities given by DL T-SDA and DL P-SDA with GA-specific transfer learning gives the highest performance on the validation data when combined with a weight of 0.7 given to P-SDA and a weight of 0.3 given to T-SDA. From Figure 5 it can indeed be seen that the fusion of the term and preterm SDAs outperforms either of them separately for the complete range of thresholds, reaching the final best overall AUC of 95.4% on the preterm test DB. The

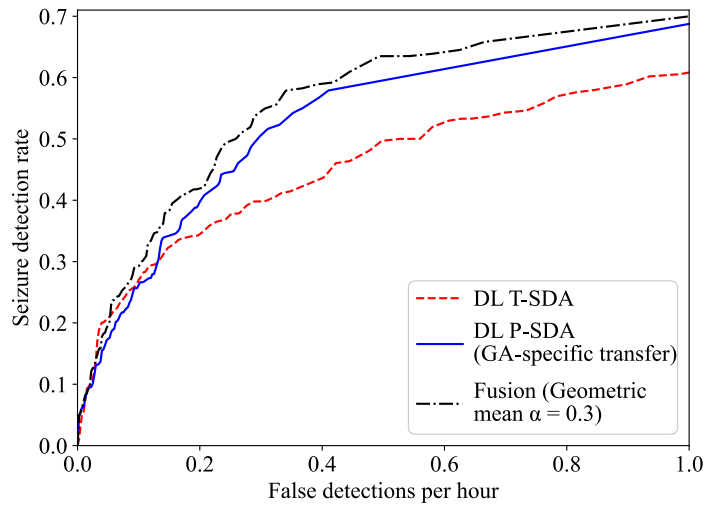

Fig. 5. The curve of seizure detection rate vs the number of false detections per hour. The fusion with geometric mean reaches an overall AUC of 95.4%.

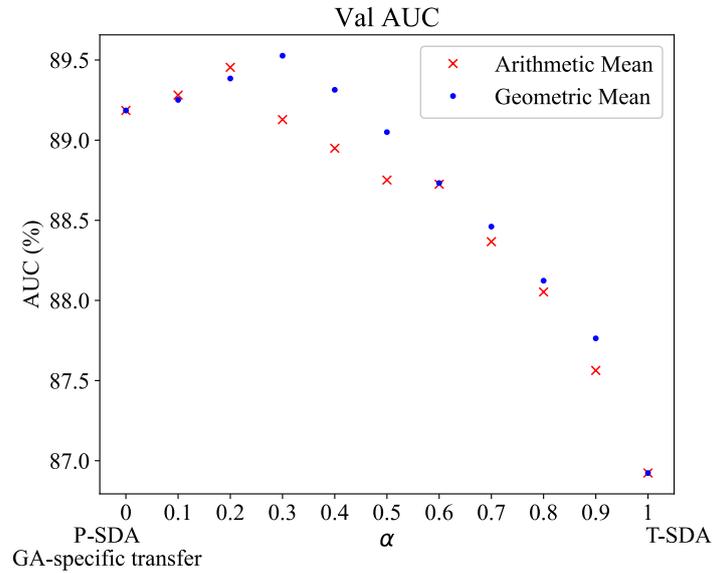

Fig. 6. The performance on all infants in the validation dataset concatenated together as a function of α. The performance reported here is for the two held-out preterms from the train DB, which were used for early stopping, for each GA-specific classifier. The leftmost point represents the DL P-SDA with GA specific transfer learning (α = 0), the rightmost point represents the DL T-SDA (α = 1). Validation AUC is reported without postprocessing and thus can be smaller than test DB performance.

Preterm EEG test DB includes infants which did not experience seizures. An AUC of 96.3 is obtained when the ensemble is tested on seizure infants only.

Figure 7 presents results of experiments where variation of classifier performance with respect to infants in the test dataset is analysed for the three DL classifiers which were studied in this work. The AUC score for each classifier was calculated on subsets of the overall test dataset; each subset was created by removing an infant from the test dataset. This resulted in 16 variations of test conditions, one corresponding to the removal of each infant in the testset. The DL T-SDA, DL P-SDA (GA-specific TL), and the Fusion (Geometric mean α = 0.3) classifiers were tested for each testing condition and the resultant distribution of AUC scores for each classifier are illustrated in a boxplot in Figure 7. These experiments indicate that the improvement in performance due to classifier training adaptions which were tested as part of this work are consistent across the cohort of infants.

## 5. Discussion

### 5.1. *Performance of T-SDAs on the preterm test DB*

Previous works have proposed several algorithms to detect seizures in full term EEG;[40,47,48] these works have achieved AUC performances in the range of 83% - 96.6% when tested on term EEG. It has been shown that both the approaches that rely on hand-crafted features and the approaches that are based on deep learning achieve reliable and robust results when trained and tested on term EEG. This work however illustrates that the performance of these approaches falls short when tested on preterm EEG to detect seizures.

The results in Table 4 indicate that the SVM T-SDA, which is based on a set of hand-crafted features and utilizes channel-specific seizure annotations during training, in particular, is not suitable for use as a preterm seizure detection tool. In contrast, the DL T-SDA was trained with multi-channel temporal EEG and did not require an engineered feature set or channel-specific seizure annotations during training. The DL T-SDA architecture is not affected by changes in the EEG montage, making it a robust classifier which performs well on unseen datasets.[19] Various ways to improve the performance of SDAs designed for term EEG are investigated in this work and assessed on the preterm test DB.

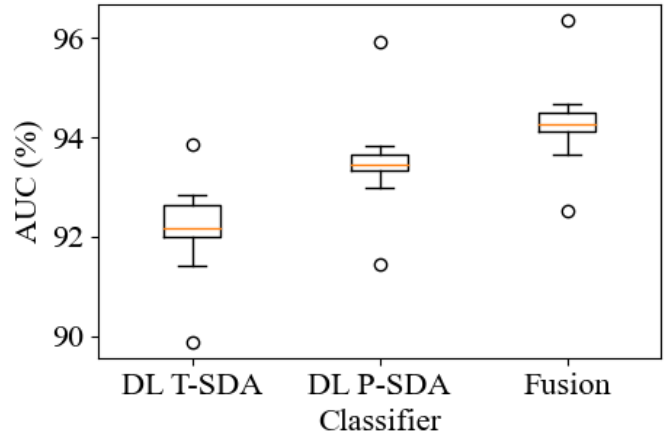

Fig. 7. Distribution of AUC scores for experiments were one infant is removed from the testset; scores are reported as % AUC with only moving average filter postprocessing applied to the probabilistic outputs (no collar and no respiration adaptation). The performances of the DL T-SDA, DL P-SDA (GA-specific TL), and the Fusion (Geometric mean α = 0.3) classifiers are illustrated in a boxplot. This analysis indicates that the developed DL P-SDA is significantly better than the DL T-SDA (p < .001). The fusion of these two classifiers is significantly better than the DL T-SDA (p < .001) and classifier fusion also results in a significantly better classifier than the DL P-SDA (p < .01).

**5.2. *Retraining on preterm data***

While the results for the T-SDAs in Table 4 show a large decrease in performance when the algorithms were tested on the preterm test DB, a marginal improvement was observed when retrained on the preterm training DB. Overall, the resultant DL P-SDA outperformed the SVM P-SDA on the preterm test DB.

A contributor to the poorer performance of the SVM SDAs in comparison to the DL SDAs is their dependence on a set of hand-crafted features which were specifically designed to capture information present in term EEG and particularly term seizures.[16,37,38] These features have been developed for term seizure detection over nearly a decade and consequently they are not optimized to capture preterm EEG specific information. Significant engineering effort would therefore be necessary to adapt the existing feature extraction method and to engineer new characteristics that would be specific to preterm seizures and preterm EEG background. The SVM T-SDA algorithm has also shown to be less effective at detecting short seizure events, which are more prevalent in the preterm EEG data, as shown in Figure 2.[49] This further highlights the mismatch between conditions for detecting term seizures and conditions for detecting preterm seizures.

A factor which affects the performance of the DL SDAs is the need for large amounts of training data; the DL T-SDA was trained on over 800 hours of multi-channel EEG, in contrast the DL P-SDA was trained on less than 24 hours of multi-channel preterm EEG. While the small improvement observed for the SVM P-SDA, when the SVM T-SDA was retrained with the preterm training DB, was mainly limited by using features designed for term infants, the small improvement that was observed in the DL P-SDA was undoubtedly limited by the size of the preterm training DB which was not sufficiently large to learn the complexity and variation in the preterm EEG with varying GA.

**5.3. *GA-specific retraining and transfer learning***

To evaluate the effect of the GA-conditional changes in preterm EEG, the performance of the SDAs was evaluated for each GA group separately and compared. Table 5 demonstrates that the performance of the SVM P-SDAs on individual GA groups varies widely depending on the GA. No clear pattern of improvement was observed when three different GA-specific SVM classifiers were developed by segmenting the preterm training DB into groups based on GA. The performance of the GA-specific SVM P-SDA is especially poor when it is tested on a concatenation of all three GA groups; this shows the variation in optimal decision thresholds across the three models - e.g. a given threshold may accurately detect seizures in Group 1 but cause many false detections in Group 3 and vice-versa. This decreases the overall AUC across concatenated data and indicates a lack of robustness in the algorithm.

The three DL SDAs were similarly built by retraining on the same GA specific data. The use of transfer learning and GA-specific models showed an improved performance for the DL P-SDA as shown in Table 5, (overall AUC improved from 93.5% to 95%). DL based algorithms require large amounts of training data. The preterm training DB is a relatively small dataset for DL algorithm development, dividing this DB into three GA specific groups would result in even smaller datasets to train each GA-specific DL P-SDA. To overcome the constraints imposed by data scarcity, three different classifiers were trained using the complete preterm training DB as explained in Section 3. Additionally, the GA specific DL P-SDA training also benefitted from the utilization of transfer learning; a deep learning technique which allows for the initialization of algorithms using pre-trained network weights borrowed from the DL T-SDA.

It is worth noting that the DL P-SDA was trained using seizure annotations that contained no information about the seizure locations. Preterm seizures are known to start and remain more focal than term seizures with less propagation.[27,28] The DL P-SDA takes multi-channel EEG as input and a single seizure or non-seizure label as target. The training dataset in this work consists of 14 channels of EEG, therefore the seizure to non-seizure ratio could be as poor as 1-to-13 for a DL P-SDA training epoch. But the ability to learn from these weakly labelled examples means that all the data in the preterm training DB were available to train the algorithm.

The development of GA-specific classifiers does come at the cost of an additional input requirement; knowledge of an infant's GA must be utilised to select the appropriate classifier. This additional classifier input constraint must be considered if GA-specific classifiers are to be employed in clinical settings.

### 5.4. *Mixture of experts*

The DL models outperform the SVM SDAs both on term and preterm test datasets, reaching an AUC of 98.5% for term and 95% for preterm as compared to 96.6% and 89.7% for SVM term and preterm SDAs, respectively. Interestingly, when retrained on preterm training DB from scratch and with transfer learning, both SVM and DL P-SDAs do not improve the performance for all groups. In fact, the best performance for Group 3 (GA > 29 weeks) is obtained with the SVM preterm model (SVM P-SDA retrained scratch), for Group 2 (26 weeks ≤ GA ≤ 29 weeks) with the deep learning term model (DL T-SDA), for Group 1 (26 weeks ≤ GA ≤ 29 weeks) and overall for all 3 groups with GA-specific transfer learning (DL P-SDA GA-specific transfer). Taking the best model for each group and using the resultant meta-classifier will inevitably improve the performance on the observed dataset. However, this will come at a cost of using the same data to select the models and assess the performance of the models which will invalidate the latter. The choice of models must be made without observing the preterm test DB performance.

The final stage of development in this work was a fusion of various models and selection of the best combination on the validation dataset which was formed from the preterm training DB. The best validation performance was obtained when combining the DL T-SDA and the GA specific DL P-SDA. These represent diverse algorithms as they utilised different training datasets. The process of selecting a fusion algorithm and a weighting coefficient, α, for the selected two algorithms was illustrated in Figure 6. This validation dataset which was previously utilised to perform early stopping was reutilised to select the fusion hyper-parameter. The combination of these two models improved the performance further to 95.4%. It is important to stress that the reported performance is not the best achievable on this dataset but the performance that is anticipated through model selection on the validation data.

While the AUC metric represents an evaluation of the epoch-based accuracy, which is an important tool for engineering development, the full picture of an algorithm's clinical utility can be seen through event-based metrics such as the seizure detection rate and the number of false detection per hour. This trade-off between event sensitivity and event specificity is shown in Figure 5. Considering a threshold of one false detection every four hours the term algorithm (DL T-SDA) detects 37.6% of seizures, the GA specific preterm algorithm with transfer learning (DL P-SDA GA-specific transfer) detects 44.6%, and the fusion of these two classifiers by geometric mean, with 0.7 and 0.3 weights for preterm and term algorithms, respectively, results in 49.7% of seizure events being detected. Importantly, this mixture of experts outperforms the best DL P-SDA GA-specific transfer for the entire range of operating points.

The inclusion of the non-seizure infants in the test DB ensures that the model is tested under a more realistic setting where a low number of false seizure alarms must be enforced. Testing this ensemble model on the subset of six seizure infants in the Preterm Test DB results in 62.4% of seizure events being detected at a threshold of one false detection every four hours, in comparison to 49.7% of seizure events detected when control infants are included.

Figure 7. further shows that the improvements seen by developing a DL GA-specific P-SDA classifiers and the fusion of two classifiers represent significant improvements which are not subject to fluctuations based on the inclusion of individual infants in the training dataset. This test indicates the robustness of the improvements which are the result of the classifier improvements in this work. All the modifications in this work were related to the DL training routine, further improvements in this area could be achieved with the incorporation of more advanced architectures. In recent works newer and more powerful architectures have been developed and applied to epilepsy detection including the Enhanced Probabilistic Neural Network, a Neural Dynamic Classification algorithm, and a Finite Element Machine[50-52]. Each novel architecture is associated with benefits for supervised pattern recognition tasks and they may prove suitable for overcoming the preterm seizure detection challenges.

### 5.5. *Limitations*

The preterm test DB and the preterm train DB utilised in this work both had temporal seizure annotations available, also known as weak labels; additionally the train DB had a subset of seizure events labelled with strong channel-specific annotations. Annotations from a single expert were available for each infant in these DBs, however it is important to highlight that there is a mis-match between the experts who provided annotations across the DBs. The variation between the expert annotators who provided labels in the training and testing DBs in this work reflects the reality of mis-matched conditions which an AI-assisted seizure detection algorithm would have to face in clinical usage.

The inter-observer agreement (IOA) for neonatal seizure annotations for experts is known to vary.[53] A true test of a neonatal SDA would be to compare the developed algorithm against multiple human annotators using the IOA metric in the test DB. This is the first time that a seizure detection algorithm has been developed for preterm infants, further research in this area would benefit from the availability of multiple sets of annotations on the test DB. The availability of multiple expert annotators in any future training DB would also be advantageous as deep learning algorithms can benefit from modelling individual labellers and learning a data-driven weighted average of these individual expert models.[54]

This work represents the first time that neonatal seizure detection algorithms have been trained or tested on preterm EEG. The datasets utilised in this work were sufficient to show the potential of deep learning for developing preterm seizure detection algorithms but further work in this area would require larger testing datasets to validate the developed classifier. However, it should be noted that this dataset of preterm EEG recordings with annotated seizures represents a relatively large dataset, considering the difficulty of recording and annotating EEG from this vulnerable preterm population.

### 6. Conclusion

This study represents the first time that an automated seizure detection system has been developed specifically for preterm EEG. This work has contributed three important messages. First the study showed that the performance of the algorithms developed for term EEG was not suitable to be used for seizure detection in preterm EEG. Second, the study showed that no substantial increase in algorithm performance was observed when the systems were simply retrained on preterm EEG. Finally, it is shown that the lengthy process of manual feature engineering could be avoided, and an accurate SDA can be developed with minimal preterm data availability. The proposed novel end-to-end deep learning approach utilizes three GA-specific models and overcomes limited data availability by 1) weighting data from all GA subgroups 2) utilising transfer learning from the term model 3) not requiring per-channel annotations and therefore learning from the more readily available weak labels. The performance obtained by combining the best performing algorithms reached an AUC score of 95.4%; while detecting nearly half of preterm neonatal seizure events at a cost of one false seizure detection every 4 hours. The results obtained using deep learning allow for its practical application in neonatal intensive care units for detection of preterm seizures. The proposed DL architecture can be improved with a larger dataset of preterm EEGs with annotated seizures without the need for channel-specific seizure annotations, thus reducing the workload on clinical staff. Future work in this area will focus on gaining a clinical understanding of the patterns in EEG which lead to false detections and missed seizure events. This future clinical analysis of EEG morphology will give engineers insights into the potential differences between SDAs developed for term and preterm EEG, and how the preterm specific patterns affect algorithm efficacy.

## 7. Conflict of interest

None of the authors have potential conflicts of interest to be disclosed.

## 8. Acknowledgements

This research was a collaborative research project (NESTED) funded by Science Foundation Ireland (INFANT-12/RC/2272). We gratefully acknowledge the financial support received from Inspiration Healthcare Ltd. in collaborating on this research. We would like to thank the families of all infants who contributed to this study in both Ireland and Italy.